\documentclass[conference,9pt]{IEEEtran}
\usepackage{graphicx}
\usepackage{amsmath}
\usepackage{amssymb}
\usepackage{adjustbox}
\DeclareMathOperator*{\argmaxA}{arg\,max} 
\newcommand{\norm}[1]{\left\lVert#1\right\rVert}
\usepackage{algorithm}
\usepackage{algorithmic}
\usepackage{array}
\usepackage{float}
\usepackage{caption}
\usepackage{subcaption}

\begin{document}
\author{Chandrashekhar Rai\IEEEmembership{Student Member, IEEE} and Debarati Sen \IEEEmembership{Senior Member, IEEE}\\
	G.S.Sanyal School of Telecommunications, IIT Kharagpur, India\\
	\small{cs.rai93@iitkgp.ac.in, debarati@gssst.iitkgp.ac.in}}

\title{Low Complexity DoA-ToA Signature Estimation for Multi-Antenna Multi-Carrier Systems}
\maketitle

\begin{abstract}
    Accurate direction of arrival (DoA) and time of arrival (ToA) estimation is an stringent requirement for several wireless systems like sonar, radar, communications, and dual-function radar communication (DFRC). Due to the use of high carrier frequency and bandwidth, most of these systems are designed with multiple antennae and subcarriers. Although the resolution is high in the large array regime, the DoA-ToA estimation accuracy of the practical on-grid estimation methods still suffers from estimation inaccuracy due to the spectral leakage effect.
    In this article, we propose DoA-ToA estimation methods for multi-antenna multi-carrier systems with an orthogonal frequency division multiplexing (OFDM) signal. In the first method, we apply discrete Fourier transform (DFT) based coarse signature estimation and propose a low complexity multistage fine-tuning for extreme enhancement in the estimation accuracy. The second method is based on compressed sensing, where we achieve the super-resolution by taking a 2D-overcomplete angle-delay dictionary than the actual number of antenna and subcarrier basis. Unlike the vectorized 1D-OMP method, we apply the low complexity 2D-OMP method on the matrix data model that makes the use of CS methods practical in the context of large array regimes. Through numerical simulations, we show that our proposed methods achieve the similar performance as that of the subspace-based 2D-MUSIC method with a significant reduction in computational complexity.
\end{abstract}


\section{Introduction}
The wireless communication signal can also be utilized for radio localization of scatters or targets present in the channel \cite{kotaru2015spotfi}. With the recent use of Millimeter-Wave (mmWave) / TeraHertz (THz) carrier frequencies \cite{tripathi2021millimeter}, there is an opportunity for mounting massive antenna numbers at both Mobile Station (MS) and Base Station (BS). With this available massive antenna size and huge Bandwidth (BW), we can achieve high-resolution DoA-ToA estimation\cite{jornet2023wireless}. However, there are certain applications where the DoA-ToA estimation and the corresponding path gain should be estimated as correctly as possible, e.g., Intelligent Reflecting Surface (IRS) \cite{liu2020deep} and angle-domain wireless transmission \cite{fan2017angle}. Moreover, the proposed estimation algorithm should be capable of accurate parameter estimation in low-measurement grid scenarios as well. For this purpose, there exist subspace-based algorithms, e.g., MUltiple SIgnal Classification \cite{schmidt1986multiple} and Estimation of Signal Parameters via Rotational Invariance Technique (ESPIRIT) \cite{roy1989esprit}. Using MUSIC, a joint DoA-ToA estimation framework for uniform circular antennas is developed in \cite{chen2017joint,chen2019low}. 2D-MUSIC-based methods are investigated for DFRC systems in \cite{bhogavalli2023estimating} and for radar in \cite{deng2024toa}. However, these techniques involve Singular Value Decomposition (SVD) and are not possible to run in real-time for high-measurement grids. 

Instead, one can use the DFT-based method for spectrum estimation, which is less complex and easily implementable on hardware. Yet, the DFT-based methods suffer from lower resolution and spectral leakage due to grid mismatch. Utilizing the scattering environment being sparse at mmWave/THz frequencies, we can still use the DFT-based method and control the spectral leakage effect via the array processing technique. The first challenge here is to identify the unknown number of scatters present in the radio environment. Once we get the correct number of scatters, we assign the peak around every scatter in Channel Impulse Response (CIR) as the coarse estimate \cite{rai2022signature,wang2018spatial}. The accuracy of this initial coarse estimate is limited to the given measurement grid. Later on, we can rotate around the coarse bin to get the exact input signature estimate. 
However, the accuracy of the fine-tuning estimate depends upon the grid size selected within a bin, and in a 2D scenario, it is computationally more expensive. Another paradigm can be the use of CS-based methods for achieving super-resolution in estimating the DoA-ToA as suggested in \cite{wu2022super}. However, in the large array regimes, the application of the vectorized-1D method is extremely complex and hence cannot be used for real-time applications.
To the best of our knowledge, there does not exist any work that discusses the low complexity and high accuracy joint DoA-ToA estimation of the multi-antenna multi-carrier systems in large array regimes.

Hence, to reduce the computational complexity while enhancing the estimation accuracy, instead of direct fine-tuning, we propose a multistage rotation method in this work. Moreover, we propose a 2D-OMP-based joint DoA-ToA estimation method that operates in large array regimes with almost the same complexity as the rotation method. We validate the efficacy of our proposed algorithms via numerical simulations and show that they give more accurate estimation while keeping the computational complexity lower, especially in high-grid measurements.

\textit{\textbf{Notations:}} We use small letter $a$ for scalar variables, small bold letter $\mathbf{a}$ for vectors, and capital bold letter $\mathbf{A}$ for matrices. $ (.)^* $ denotes the conjugate, $(.)^T$ denotes the transpose, and $ (.)^H $ denotes the conjugate and transpose operation. $\lfloor x \rceil
$ denotes the closest integer number to a real number $x$.$diag\lbrace \mathbf{a}\rbrace$ denotes the diagonal matrix with elements of $\mathbf{a}$ on its principal diagonal.

\section{Signal Model}

Let us have $R$ number of antenna elements arranged in a 
$d$-equispaced Uniform Linear Array (ULA) geometry. The transmitted baseband signal $x(t)$ reflected from the
$q^{th}$ point scatter received at an $r^{th}$ antenna element can be written as
\par\noindent\small
\begin{equation}
    y_r(t) = \sum_{q=0}^{Q} \alpha_{q}^{'} x(t-\tau_{q}^{'}) e^{-j2\pi f_c \tau_{q}^{'}} +e_r(t)
    \label{eq:received_signal}
\end{equation}
\normalsize
where $e_r(t)$ is the complex Additive White Gaussian Noise (AWGN) at the $r^{th}$ receiver, $\alpha_{q}^{'}$ is the complex fading coefficient of $q^{th}$ path, $\tau_{q}^{'}$ is the overall delay of $q^{th}$ path where $\tau_{q}^{'} = \tau_{0,q}+\tau_{r,q}$, with $\tau_{0,q} = r_q/c$ being the delay due to radial distance up to the $0^{th}$ (reference) antenna and $\tau_{r,q} = r d sin \theta_{q}^{'} /c$  being the spatial delay at $r^{th}$ element across the aperture length.

We can write the equivalent  space-delay CIR from \eqref{eq:received_signal} as
\begin{equation}
    h_r(\tau) = \sum_{q=0}^{Q} \alpha_{q}^{'} \delta(\tau-\tau_{q}^{'}) e^{-j2\pi f_c \tau_{q}^{'}} +e_r(\tau).
    \label{eq:continuous_delay_space}
\end{equation}

We assume the spatial narrowband assumption across the aperture of ULA, i.e., $x(t-\tau_{0,q}) \approx x(t-\tau_{R,q})$ and define $\alpha_q \triangleq \alpha_{q}^{'} e^{-j 2 \pi f_c \tau_{0,q}}$. Hence, we can rewrite \eqref{eq:continuous_delay_space} as

\begin{equation}
    h_r(\tau) = \sum_{q=0}^{Q} \alpha_q \delta(\tau-\tau_{0,q}) e^{-j2\pi f_c \tau_{r,q}} +e_r(\tau).
    \label{eq:continuous_delay_space_1}
\end{equation}

We take the Continuous Time Fourier Transform (CTFT) to get the space-frequency response as
\begin{equation}
    H_r(f) = \sum_{q=0}^{Q} \alpha_q e^{-j2\pi f \tau_{0,q}} e^{-j2\pi f_c rd sin\theta_{q}^{'}/c} +e_r(f).
    \label{eq:continuous_frequency_space}
\end{equation}

For the OFDM waveform, the system BW $f_s$ is divided in $S$ number of subcarriers with the inter-carrier spacing $\delta_f \triangleq f_s/S$. We write the sampled space-frequency response as
\begin{equation}
    H_r(s) = \sum_{q=0}^{Q} \alpha_q e^{-j2\pi s \delta_f \tau_{0,q}} e^{-j2\pi f_c rd sin\theta_{q}^{'}/c} +e_r(n).
    \label{eq:discrete_frequency_space}
\end{equation}
By defining normalized ToA $\tau_q \triangleq \delta_f \tau_{0,q}$ and normalized DoA $\theta_q \triangleq f_cdsin \theta_{q}^{'}$, we can reformulate \eqref{eq:discrete_frequency_space} as
\begin{equation}
    H_r(s) = \sum_{q=0}^{Q} \alpha_q e^{-j2\pi s \tau_q} e^{-j2\pi r \theta_q} +e_r(n).
    \label{eq:discrete_frequency_space_1}
\end{equation}

Further, we define the delay and the angle steering vectors as $\mathbf{b}(\tau) \triangleq [1, e^{-j2\pi \tau}, \cdots, e^{-j2\pi (S-1)\tau}]^T$ and $\mathbf{a}(\theta) \triangleq [1, e^{-j2\pi \theta}, \cdots, e^{-j2\pi (R-1)\theta}]^T$. Now we write the equivalent matrix form of the \eqref{eq:discrete_frequency_space_1} $\forall s \in \lbrace
0,\cdots,S-1 \rbrace,r \in \lbrace 0,\cdots,R-1 \rbrace$ as
\begin{equation}
    \mathbf{H} = \sum_{q=0}^{Q} \alpha_q \mathbf{a}(\theta_q)  \mathbf{b}^T(\tau_q) + \mathbf{E}.
    \label{eq:discrete_frequency_space_matrix}
\end{equation}

In this work, our target is to estimate the signature of the wireless radio environment defined as $\mathcal{S} \triangleq \lbrace \alpha_q, \tau_q, \theta_q, Q \rbrace$. By defining $\mathbf{X} \triangleq diag\lbrace \alpha_1 , \hdots , \alpha_Q \rbrace$, $\mathbf{A} \triangleq [\mathbf{a}(\tau_1),\hdots, \mathbf{a}(\tau_Q)]$, and $\mathbf{B} \triangleq [\mathbf{b}(\theta_1), \hdots , \mathbf{b}(\theta_Q)]$, we can write \eqref{eq:discrete_frequency_space_matrix} equivalently as
\begin{equation}
    \mathbf{H} = \mathbf{AXB^T}+ \mathbf{E}.
    \label{eq:discrete_frequency_space_matrix1}
\end{equation}

\section{Proposed Signature Estimation Methods}
We can observe from \eqref{eq:discrete_frequency_space_1} that the signal model closely resembles to a mixture of 2-D complex sinusoids, and hence it is mainly the frequency estimation problem. We take the normalized 2D IDFT of \eqref{eq:discrete_frequency_space_1} and get the sampled delay-angle channel response but with the limited resolution. The 2D-IDFT can be written mathematically as

\begin{adjustbox}{width=1.0\linewidth}
\begin{minipage}{\linewidth}
\begin{equation}
\begin{aligned}
  G(i,j) &= \sum_{q=0}^{Q}\frac{\alpha_q}{\sqrt{RS}} e^{-j\pi(\theta_q-\frac{i}{R})} e^{-j\pi(\tau_q-\frac{j}{S})} \\
  & \quad  \underbrace{\frac{sin \pi R (\theta_q-\frac{i}{R})}{sin \pi (\theta_q-\frac{i}{R})}}_{D_R(\theta_q-\frac{i}{R})} \underbrace{\frac{sin \pi S (\tau_q-\frac{j}{S})}{sin \pi (\tau_q-\frac{j}{S})}}_{D_S(\tau_q-\frac{j}{S})}. \\
  \label{eq:narrowband_spectral_leakage}
\end{aligned}
\end{equation}
\end{minipage}
\end{adjustbox}

In the asymptotic case $lim_{n \rightarrow \infty} \frac{1}{\sqrt{N}} |D_N(x)| = \sqrt{N}\delta(x)$, and hence when $R, S \rightarrow \infty$, we can have the exact signature estimate by picking the peaks of the angle-delay magnitude spectrum

\begin{equation}
   |G(i,j)| = \sum_{q=0}^{Q} \alpha_q \delta\left(\theta_q-\frac{i}{R}\right
   ) \delta\left(\tau_q-\frac{j}{S}\right).
\end{equation}

However, in practice, it is not possible to have the infinite antennae and subcarriers. Hence, there will be significant power leakage to the adjacent bins, and the estimation accuracy will be compromised (even worsens when the measurement grid size is low). To handle the leakage effect and to get the accurate signature estimate we propose two low complexity methods as described in subsequent sections.

\subsection{Method-1: 2D Rotation}
\subsubsection{Direct Rotation} We can identify that the power around each scatter leaks when the input signature $(\theta_q,\tau_q)$ is not matched to the grid $(\frac{i}{R},\frac{j}{S})$. The leakage can be handled around the coarsely estimated bin $(i_q,j_q)$ via the rotation technique. We rotate by a fractional amount $(\theta_q^r, \tau_q^r)$ such that all the path power is concentrated to a bin. We define the rotation matrices as $\mathbf{R} ({\theta_q^r}) \triangleq diag \lbrace 1, e^{j 2 \pi \theta^r_q}, \cdots, e^{j 2 \pi (R-1)\theta^r_q}\rbrace$, and $\mathbf{R}(\tau_r^q) \triangleq diag \lbrace 1, e^{j 2 \pi \tau^r_q}, \cdots, e^{j 2 \pi (R-1)\tau^r_q}\rbrace$. We can write the rotated space-frequency channel matrix as

\begin{equation}
    \mathbf{H}^r =  \mathbf{R}_{\theta} \mathbf{H} \mathbf{R}_{\tau}.
    \label{eq:rotated_cir}
\end{equation}

We take the IDFT of \eqref{eq:rotated_cir} and get the corresponding rotated angle-delay response as 

\begin{adjustbox}{width=1.0\linewidth}
\begin{minipage}{\linewidth}
\begin{equation}
\begin{aligned}
  G(i,j) &= \sum_{q=0}^{Q}\frac{\alpha_q}{\sqrt{RS}} e^{-j\pi(\theta_q+\theta_q^r-\frac{i}{R})} e^{-j\pi(\tau_q+\tau_q^r-\frac{j}{S})} \\
  & \quad  \frac{sin \pi R (\theta_q+\theta_q^r-\frac{i}{R})}{sin \pi (\theta_q+\theta_q^r-\frac{i}{R})} \frac{sin \pi S (\tau_q+\tau_q^r-\frac{j}{S})}{sin \pi (\tau_q+\tau_q^r-\frac{j}{S})}. \\
  \label{eq:rotated_delay_angle}
\end{aligned}
\end{equation}
\end{minipage}
\end{adjustbox}

We can find the optimal rotation pair $(\theta_q^r,\tau_q^r)$ such that the entire power is concentrated  within a bin resolution. In fact, there can be an infinite point in the search grid, though we limit the search to the finite grid points around the coarse bin. Therefore, the accuracy provided by the rotation algorithm is inversely proportional to the direct grid search around the coarse bins. Hence, to reduce the computational complexity, instead of searching with a fine grid, we divide the grid search into the multiple stages. First, let us see the direct search methodology. We see that the search grid for the $q^{th}$ path is $(\theta_q^r,\tau_q^r) \in (-\frac{i_q}{2R},\frac{i_q}{2R}) \times (-\frac{j_q}{2S},\frac{j_q}{2S})$ around the $q^{th}$ coarse bin. Let us divide the search grid with $N_{\theta}^r, N_{\tau}^r$ fine levels, i.e.,  $(\theta_q^r,\tau_q^r) \in (S_{\theta_q},S_{\tau_q}) =  \big(-\frac{i_q}{2R}:\frac{1}{N_{\theta}^r}:\frac{i_q}{2R}\big) \times \big(-\frac{j_q}{2S}:\frac{1}{N_{\tau}^r}:\frac{j_q}{2S}\big)$. We can estimate the fine-tuned normalized DoA-ToA by picking the peak in the rotated spectrum as

\begin{equation}
    \begin{aligned}
    (\hat{\theta}_q^r,\hat{\tau}_q^r) 
     = \argmaxA_{(\theta_q^r,\tau_q^r)\in (S_{\theta_q},S_{\tau_q})} \norm{\mathbf{f}^{H}_{i_q,:} \mathbf{R}(\theta^r_q) \mathbf{H} \mathbf{R}(\tau^r_q) \mathbf{f}_{j_q,:}}_2^2.\\
    \end{aligned}
\end{equation}
\begin{equation}
    \hat{\theta}_q = \frac{i_q}{R}+\hat{\theta}^r_q, \quad \hat{\tau}_q = \frac{j_q}{S}+\hat{\tau}^r_q.
\end{equation}

\begin{equation}
    \hat{\alpha}_q =  [\mathbf{f}^{H}_{i_q,:} \mathbf{R}_{\hat{\theta}_q} \mathbf{H} \mathbf{R}_{\hat{\tau}_q}\mathbf{f}_{j_q,:}].
\end{equation}

In this way, we need to search with $N_{\theta}^rN_{\tau}^r$ points around every coarse bin. However, if the measurement grid is already less, we need to put more points for the fine-tuning, which makes the direct rotation time-consuming.
\subsubsection{Multistage Rotation}
\begin{figure}[htbp]
    \centering
    \includegraphics[width = 0.95\linewidth]{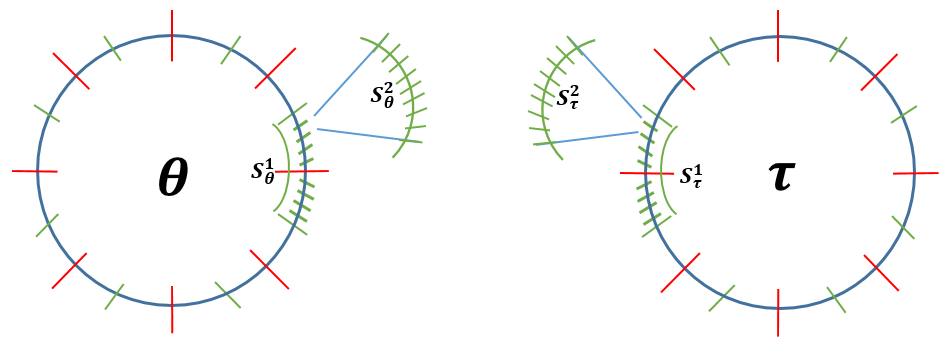}
    \caption{Multistep rotation method}
    \label{fig:rotation}
\end{figure}

We apply the approach similar to the binary search method as illustrated in Fig. \ref{fig:rotation}. At first, we divide the grid with $(N_{\theta}^{r_1}, N_{\tau}^{r_1})$ points and search for the maximum in $N_{\theta}^{r_1}, N_{\tau}^{r_1}$ (quite less fine levels) $(\theta_q^{r_1},\tau_q^{r_1}) \in (S^1_{\theta_q},S^1_{\tau_q}) =  \big(-\frac{i_q}{2R}:\frac{1}{N_{\theta}^{r_1}}:\frac{i_q}{2R}\big) \times \big(-\frac{j_q}{2S}:\frac{1}{N_{\tau}^{r_1}}:\frac{j_q}{2S}\big)$. We find the peak in the first stage grid $(i_q^1,j_q^1)$, and we further divide around this with $N_{\theta}^{r_2}, N_{\tau}^{r_2}$ points as  $(\theta_q^{r_2},\tau_q^{r_2}) \in (S^2_{\theta_q}, S^2_{\tau_q}) =  \big(-\frac{i^1_q}{2R}:\frac{1}{N_{\theta}^{r_2}}:\frac{i^1_q}{2R}\big) \times \big(-\frac{j^1_q}{2S}:\frac{1}{N_{\tau}^{r_2}}:\frac{j^1_q}{2S}\big)$. 

We can write the divided search in two stages mathematically as
\begin{equation}
\begin{aligned}
    (\theta^{r_1}_{q},\tau^{r_1}_{q}) & = \argmaxA_{(\theta_q^{r_1},\tau_q^{r_1}) \in (S_{\theta_q}^{1}, S_{\tau_q}^{1})} \norm{\mathbf{f}^{H}_{i_q,:} 
 \mathbf{R_{\theta^1_q}} \mathbf{H} \mathbf{R_{\tau^1_q}}  \mathbf{f}_{j_q,:}}_{2}^{2} \\
    (\theta^{r_2}_{q},\tau^{r_2}_{q}) & = \argmaxA_{(\theta_q^{r_2},\tau_q^{r_2}) \in (S_{\theta_q}^{2}, S_{\tau_q}^{2})} \norm{\mathbf{f}^{H}_{i_q,:} 
 \mathbf{R_{\theta^2_q}} \mathbf{H}  \mathbf{R_{\tau^2_q}}  \mathbf{f}_{j_q,:}}_{2}^{2} .
    \label{eq:rotation_first}
\end{aligned}
\end{equation}

Finally, we can estimate the fine-tuned signature after using the two-stage rotation method as

\begin{equation}
    \hat{\theta}_q = \frac{i_q}{R}+\hat{\theta}^{r_2}_q, \quad \hat{\tau}_q = \frac{j_q}{S}+\hat{\tau}^{r_2}_q.
\end{equation}
\begin{equation}
    \hat{\alpha}_q = [\mathbf{f}^{H}_{i_q,:} \mathbf{R}_{\hat{\theta}^2_q} \mathbf{H} \mathbf{R}_{\hat{\tau}^2_q} \mathbf{f}_{j_q,:}].
\end{equation}

We give the concrete mathematical signature estimation steps in Algorithm \ref{algorithm:algorithm1}.

\begin{algorithm}
\caption{Multistep Rotation Signature Estimation}
\label{algorithm:algorithm1}
    \begin{algorithmic}[1]
        \REQUIRE Measured Input CIR Matrix $\mathbf{H}$, Rotation Count-$N_{\theta}^{r_1},N_{\tau}^{r_1},N_{\theta}^{r_2}, N_{\tau}^{r_2}$ 
        \ENSURE Signature Estimate $ \lbrace \hat{Q}, \hat{\theta}_q, \hat{\tau}_q, \hat{\alpha}_q \rbrace $ 
        \STATE Calculate the initial delay-angle CIR via 2D IDFT $\mathbf{G}(i,j)$
        \STATE Find the 2D-peaks in $\mathbf{G}$ and coarse estimate $\lbrace \hat{Q},i_q,j_q \rbrace$
        \FOR {$q = 1: \hat{Q}$ }
              
              \FOR{$\frac{-i_q}{2R}: \frac{1}{N_{\theta}^{r_1}} :\frac{i_q}{2R}, \quad \frac{-j_q}{2S}: \frac{1}{N_{\tau}^{r_1}} :\frac{j_q}{2S}$}
                    \STATE $  (\hat{\theta}^{r_1}_{q},\hat{\tau}^{r_1}_{q}) =  \underset{(\theta_q^{r_1},\tau_q^{r_1})} {\mathrm{argmax}}\norm{\mathbf{f}^{H}_{i_q,:}\mathbf{R(\theta^{r_1}_q)}\mathbf{H} \mathbf{R(\tau^{r_1}_q)}\mathbf{f}_{j_q,:}}_2^2$
                    \STATE Find for maximum $(i^1_q,j^1_q) \in S_{\theta_q}^1 \times  S_{\tau_q}^1$
              \ENDFOR
              \FOR{$\frac{-i^1_q}{2R}: \frac{1}{N_{\theta}^{r_2}} :\frac{i^1_q}{2R}, \quad \frac{-j^1_q}{2S}: \frac{1}{N_{\tau}^{r_2}} :\frac{j^1_q}{2S}$}
                    \STATE 
                    $  (\hat{\theta}^{r_2}_{q},\hat{\tau}^{r_2}_{q}) =  \underset{(\theta_q^{r_2},\tau_q^{r_2})} {\mathrm{argmax}} \norm{\mathbf{f}^{H}_{i_q,:}\mathbf{R(\theta^{r_2}_q)}\mathbf{H} \mathbf{R(\tau^{r_2}_q)}\mathbf{f}_{j_q,:}}_2^2$
              \ENDFOR
              \STATE $\hat{\theta}_q = \frac{i_q}{R}+\hat{\theta}^{r_2}_q, \quad \hat{\tau}_q = \frac{j_q}{S}+\hat{\tau}^{r_2}_q$
              \STATE $\hat{\alpha}_q = \norm{\mathbf{f}^{H}_{i_q,:}\mathbf{R(\hat{\theta}^{r_2}_{q})}\mathbf{H}\mathbf{R(\hat{\tau}^{r_2}_{q})}\mathbf{f}_{j_q,:}}_2^2$
        \ENDFOR
    \end{algorithmic}
\end{algorithm}

\subsection{Method-2 OMP-based Recovery} 

We can exploit the sparsity of the radio environment and use the CS-based methods to get the signature estimation. To this end, we consider $P_{\theta}$ and $P_{\tau}$ grid points for angle and delay, respectively while both $P_{\theta},P_{\tau} \gg Q$ and negligible discretization error. Using these grid points, we construct the overcomplete  dictionaries $\mathbf{\Tilde{A}} = [\mathbf{a}(\phi_1), \hdots, \mathbf{a}(\phi_{P_\theta})]_{R\times P_{\theta}}$ and $\mathbf{\Tilde{B}} = [\mathbf{b}(\psi_1), \hdots, \mathbf{b}(\psi_{P_\theta})]_{S\times P_{\tau}}$ for angle and delay domain respectively. Equivalently, we can represent \eqref{eq:discrete_frequency_space_matrix1} as 

\begin{equation}
    \mathbf{H} = \mathbf{\Tilde{A}\Tilde{X}\Tilde{B}^T}+ \mathbf{E}
    \label{eq:discrete_frequency_space_matrix2}
\end{equation}

where, sparse $\Tilde{X}$ contains the coefficients $\lbrace \Tilde{\alpha}_q \rbrace$ and scatter DoA-ToA information. Specifically, a non-zero element in sparse $\mathbf{\Tilde{X}}$ represents a scatter with its normalized DoA and ToA being equal to that of the corresponding 2D grid point. In \eqref{eq:discrete_frequency_space_matrix2}, the recovery is dependent upon the choice of grid points.
The 2D sparse recovery problem \eqref{eq:discrete_frequency_space_matrix2} can be solved using either 1D vectorized CS or direct 2D-OMP methods. 
\subsubsection{1D-OMP}
For the vectorized method, we denote $\mathbf{h}=\textrm{vec}(\mathbf{H})$, $\mathbf{\Tilde{x}}=\textrm{vec}(\mathbf{\Tilde{X}})$, $\mathbf{e}=\textrm{vec}(\mathbf{E})$ and $\mathbf{\Tilde{D}}=\mathbf{\Tilde{B}}\otimes\mathbf{\Tilde{A}}$. The equivalent 1D sparse recovery problem of \eqref{eq:discrete_frequency_space_matrix2} is then 
\par\noindent\small
\begin{align}
    \mathbf{h}=\mathbf{\Tilde{D}}\mathbf{\Tilde{x}}+\mathbf{e}.
    \label{eq_discrete_space_frequency_vectorized}
\end{align}
\normalsize

We can apply 1D-OMP method to recover the $\Tilde{\mathbf{x}}$ from \eqref{eq_discrete_space_frequency_vectorized}. However, in the large array regimes, the vectorized model in \eqref{eq_discrete_space_frequency_vectorized} causes prohibitive computational complexity.

\subsubsection{2D-OMP} We apply the 2D-OMP method to recover the sparse $\Tilde{\mathbf{X}}$ matrix directly from \eqref{eq:discrete_frequency_space_matrix2}. Particularly, we adapt the 2D-OMP method suggested in \cite{fang20122d} to complex-valued fields and rectangular observation model
in order to work for our multi-antenna multi-carrier system. The proposed 2D-OMP methodology reduces the computational complexity significantly as will be seen in Section \ref{sec:results}.
\begin{figure}
\centering
\begin{subfigure}{.5\linewidth}
  \centering
  \includegraphics[width=.97\linewidth]{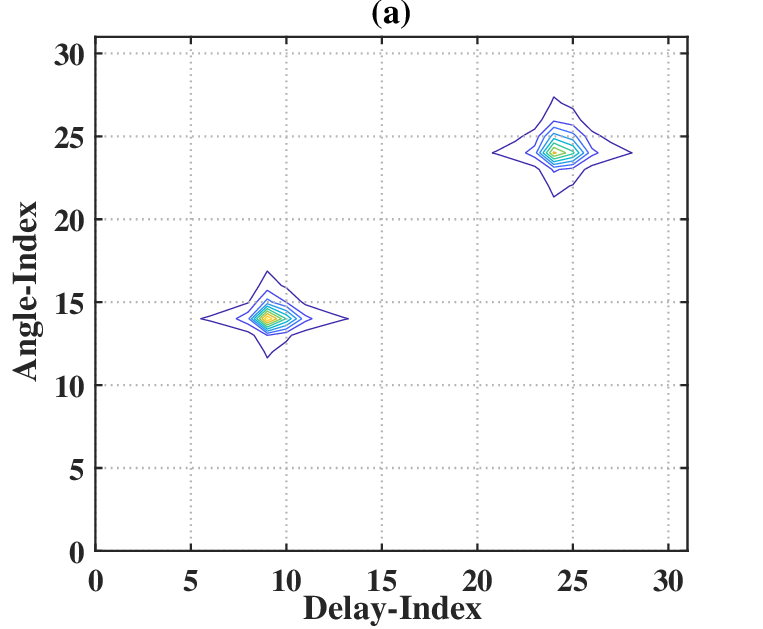}
\end{subfigure}%
\begin{subfigure}{.5\linewidth}
  \centering
  \includegraphics[width=.97\linewidth]{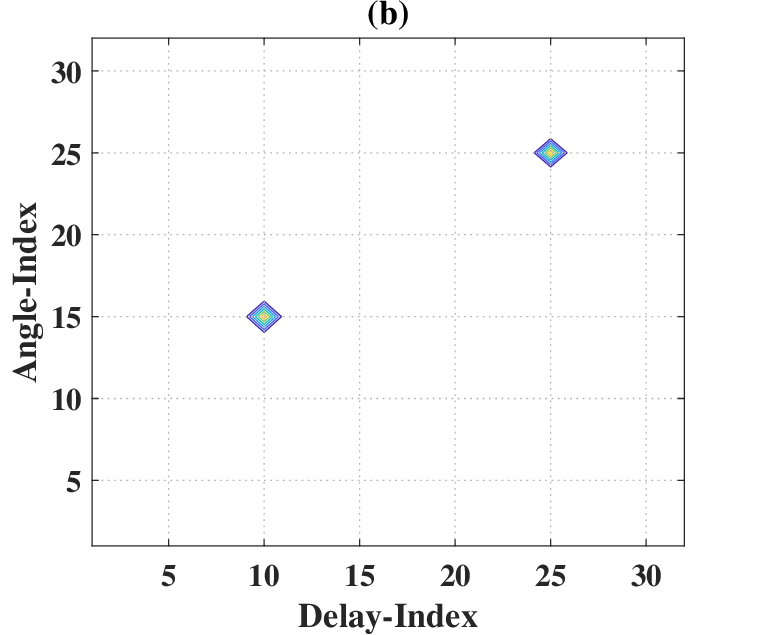}
\end{subfigure}
\caption{Delay-Angle CIR for (a) Direct DFT (b) Two-stage Rotation.}
\label{fig:rotation_correction}
\end{figure}
\subsection{Computational Complexity} In the direct rotation method, we have to do an exhaustive search in the entire $N^r_{\theta}, N^r_{\tau}$ grid points in order to get the fine-tuned estimations. The complexity is dominated by the fine-tuning step and with  $N^r_{\theta}, N^r_{\tau}$ grid-points search requires $\mathcal{O}(QRSN^r_{\theta}N^r_{\tau})$. In multistep method, if we divide $N^r_{\theta} = N^{r_1}_{\theta}N^{r_2}_{\theta}$ and $N^r_{\tau} = N^{r_1}_{\tau}N^{r_2}_{\tau}$,  we have to search only for $log_{(N^{r_1}_{\theta})}(N^{r}_{\theta}), log_{(N^{r_1}_{\tau})}(N^{r}_{\tau})$ grid points. Hence the complexity will be $\mathcal{O}(QRS \mathbf{log} (N^{r}_{\theta} N^{r}_{\tau}))$. In 1D-OMP method the complexity is mainly dominated by the projection step and is $\mathcal{O}(QSR P_{\theta} P_{\tau})$. Whereas, for $S \geq R$ the complexity of 2D OMP method is $\mathcal{O}(QSP_{\theta}P_{\tau})$. Moreover, except for the spectrum search, the complexity of the subspace-based methods is $\mathcal{O}(R^3S^3)$. 
\begin{figure*}
\centering
\begin{subfigure}{0.32\textwidth}
    \includegraphics[width=\textwidth]{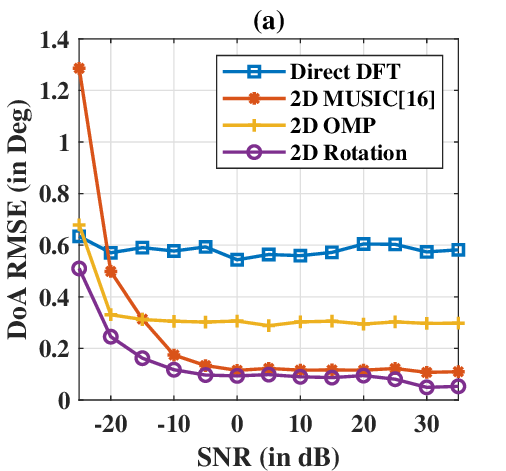}
\end{subfigure}
\hfill
\begin{subfigure}{0.32\textwidth}
    \includegraphics[width=\textwidth]{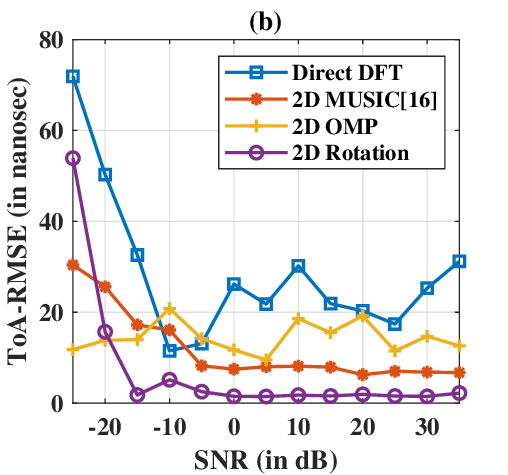}
\end{subfigure}
\hfill
\begin{subfigure}{0.32\textwidth}
    \includegraphics[width=\textwidth]{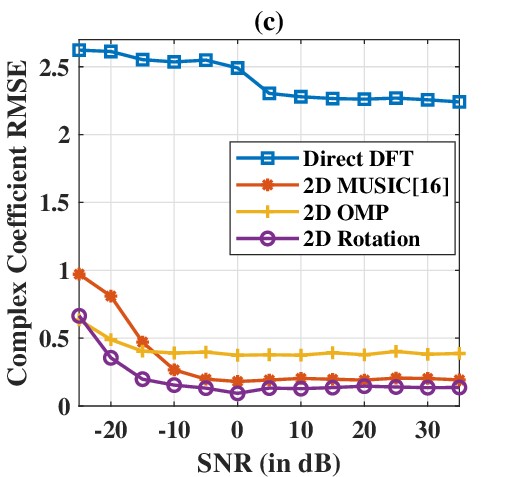}
\end{subfigure}
\caption{RMSE in (a) DoA (b) ToA, and (c) Complex coefficients for proposed, DFT and MUSIC methods.}
\label{fig:rmse}
\end{figure*}

\section{Results} 
\label{sec:results}
We first show the spectral leakage effect and its control via the rotation technique. Let there be two scatters present in the channel at normalized delay-angle being $(0.4766,0.3241)$ and $(0.7922,0.7947)$ with a fixed complex path coefficient of $0.5+0.5i$ for each path. For a measurement grid of $Nx, Ny = 32$, the input signature is $(15.25/Nx,10.37/Ny)$ and $(25.35/Nx,25.43/Ny)$ for which the spectrally leaked angle-delay CIR is shown in Fig. \ref{fig:rotation_correction}(a). Now, to compare the performance, we implement the direct rotation and multistep rotation techniques with different rotation grid sizes. With $N^r_{\theta},N^r_{\tau} = 11$, we get the estimate of path-1 being $(\hat{\theta}_1=0.4781,\hat{\tau}_1=0.3250,\hat{\alpha}_1=0.6024+0.3627i)$ and $(\hat{\theta}_2=0.7906,\hat{\tau}_2=0.7937,\hat{\alpha}_2=0.3627+0.6024i)$. On the other hand, if we take $N^r_{\theta}, N^r_{\tau} = 101$, then we get the exact signature estimate as that of the input. However, we get the same estimate with 
$N^{r_1}_{\theta}, N^{r_1}_{\tau} = 11$ and $N^{r_2}_{\theta}, N^{r_2}_{\tau} = 11$ with a comparatively less run time complexity. The identification of the correct normalized delay-angle after two-stage rotation is depicted in Fig. \ref{fig:rotation_correction}(b).

For simulations, we consider the $f_c = 73$ GHz, $f_s = 1$ GHz, $N_x = N_y = 64$ unless stated otherwise.
We take $Q=5$ targets with uniformly drawn delays from $[0,333ns]$ and uniformly drawn angles from $[10^{\circ}, 80^{\circ}]$. The paths were assumed with unity gain and uniformly distributed phases from $[0,2\pi]$. For two-stage rotation-based method, we take $N^{r_1}_{\theta}, N^{r_1}_{\tau} = 11$ and $N^{r_2}_{\theta}, N^{r_2}_{\tau} = 5$. Morover, for CS-based joint delay-angle estimation, the normalized delay and angle grid are constructed with 200 uniform points. In order to compare our proposed methods, we implement the 2D MUSIC method with smoothing as described in \cite{belfiori20122d}. We define a detected scatter as hit if the estimated ToA and DoA are within resolution limits and false otherwise. Throughout the simulations the threshold was set to guarantee the constant false alarm and the hit rate was observed.The performance of the estimated DoAs-ToAs were measured in terms of root mean squared error only for the hit case.

\begin{figure}
\centering
\begin{subfigure}{0.49\linewidth}
    \includegraphics[width=0.97\linewidth]{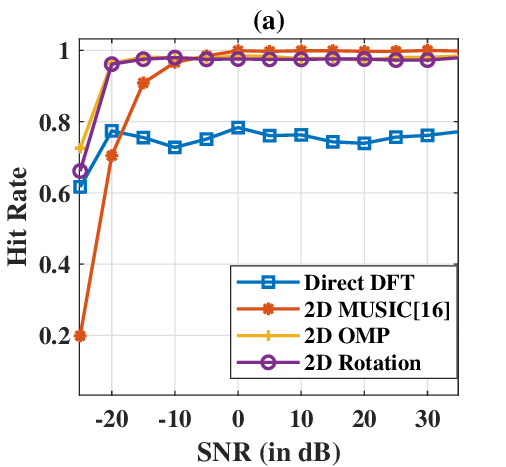}
\end{subfigure}
\hfill
\begin{subfigure}{0.49\linewidth}
    \includegraphics[width=0.97\linewidth]{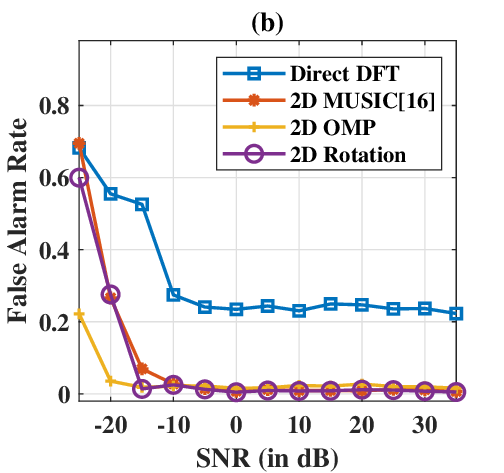}
\end{subfigure}
\caption{Average (a) false-alarm rate, and (b) hit rate at different SNRs for the proposed methods, compared with DFT and MUSIC methods.}
\label{fig:rates OMP DFT MUSIC}
\end{figure}

We show the hit rate and false alarm rate for varying SNR range from -25 dB to 35 dB in Fig. \ref{fig:rates OMP DFT MUSIC} and the corresponding RMSEs in the DoA, ToA, and complex path coefficient are shown in Fig. \ref{fig:rmse}.
Based on the definition of hit rate and false rate, we see that the DoA-ToA estimate values of the Direct DFT are limited by the measurement grid and the the corresponding hit rate does not improve with the SNR. Whereas, our proposed method achieves the same hit rate and false rate to that of the 2D-MUSIC method but with the reduced complexity. On our simulation platform (intel core i7 $10^{th}$ gen), the run time complexity for the MUSIC method is around 1 second, whereas for 2D-OMP method it is 75 milliseconds. More importantly, with our proposed two-stage rotation method, it is just 15 milliseconds. 
We show the run time complexity of the proposed methods with different measurement grid and number of scatters in Table \ref{table:complexity}. We observe that for 2D-MUSIC and 2D-Rotation method the complexity does not depend upon the number of targets, whereas for the 2D-OMP, it takes more time w.r.t. the number of targets. Among all the methods, our proposed two-stage rotation method gives the least complexity, even for higher grid size. On the other hand, the algorithm delay is quite high for subspace methods with grid size.

From Fig. \ref{fig:rmse}(a), we observe that the RMSE in DoA of the proposed two-stage rotation is the mimimum. On the other hand, the RMSE of proposed 2D-OMP method is lower bounded by the 2D-MUSIC method which is due to the limited grid size of the 2D-OMP method. Morover, one should note that the RMSE of DoA and normalized DoA has a non-linear relation. Similarly, RMSE in the ToA is the least for the proposed two-stage rotation and the worst for the direct DFT method as depicted in Fig. \ref{fig:rmse}(b). Finally, we compare the RMSE of the complex path coefficient in Fig. \ref{fig:rmse}(c) and we see that the RMSE for DFT is quite high whereas the RMSE of our proposed two-stage rotation method is below the 2D MUSIC\textit{}.
\begin{table}
\caption{Run-Time Complexity (in seconds)}
\label{table:complexity}
\resizebox{\columnwidth}{!}{%
\begin{tabular}{|c|ccc|ccc|}
\hline
            & \multicolumn{3}{c|}{Q = 5}                                    & \multicolumn{3}{c|}{Q =10}                                    \\ \hline
Method &
  \multicolumn{1}{c|}{$64\times64$} &
  \multicolumn{1}{c|}{$128\times128$} &
  $256\times256$ &
  \multicolumn{1}{c|}{$64\times64$} &
  \multicolumn{1}{c|}{$128\times128$} &
  $256\times256$ \\ \hline
2D-Rotation & \multicolumn{1}{c|}{0.015} & \multicolumn{1}{c|}{0.031} & 0.13  & \multicolumn{1}{c|}{0.015} & \multicolumn{1}{c|}{0.037} & 0.128  \\ \hline
2D-OMP      & \multicolumn{1}{c|}{0.07} & \multicolumn{1}{c|}{0.09} & 0.16  & \multicolumn{1}{c|}{0.1}  & \multicolumn{1}{c|}{0.13} & 0.23  \\ \hline
2D-MUSIC \cite{belfiori20122d}   & \multicolumn{1}{c|}{0.88} & \multicolumn{1}{c|}{4.65} & 18.85 & \multicolumn{1}{c|}{0.89} & \multicolumn{1}{c|}{4.71} & 20.57 \\ \hline
\end{tabular}%
}
\end{table}

\section{Conclusion} 
In this study, we propose algorithms for estimating the DoA-ToA in a multi-antenna multi-carrier wireless system using array processing technique and compressive sensing. We propose a low-complexity rotation-based signature estimation methodology by dividing the fine-tuning into multistage grid search while simultaneously improving in accuracy. The proposed novel fine-tuning approach in this work results in a logarithmic reduction in the computational complexity w.r.t. grid search. Furthermore, the proposed 2D-OMP method offers lower computational complexity for large array scenarios compared to the prohibitively high complexity associated with the vectorized 1D-OMP. Through various numerical simulations, we have demonstrated that our proposed methods achieve performance comparable to that of subspace-based approaches but with lower complexity.

\ifCLASSOPTIONcaptionsoff
\newpage
\fi
\bibliographystyle{IEEEtran}
\bibliography{Biblli1}
\end{document}